\documentclass[useAMS,usenatbib]{mn2e}
\usepackage{aas_macros,graphicx,times,multirow,amsmath}
\usepackage[]{color}

\title[Discovery of pulsations in 1RXS\,J225352.8+624354]{Discovery of 47-s pulsations in the X-ray source 1RXS\,J225352.8+624354}
\author[P.~Esposito et al.] {P.~Esposito,$^{1}$\thanks{E-mail: paoloesp@iasf-milano.inaf.it} G.~L.~Israel,$^{2}$ L.~Sidoli,$^{1}$ E.~Mason,$^3$ 
G.~A.~Rodr\'iguez~Castillo,$^{2,4}$ \newauthor J.~P. ~Halpern,$^5$ A.~Moretti$^6$ and D.~G\"otz$^7$\smallskip\\
$^1$Istituto di Astrofisica Spaziale e Fisica Cosmica - Milano, INAF, via E. Bassini 15, I-20133 Milano, Italy\\
$^2$Osservatorio Astronomico di Roma, INAF, via Frascati 33, I-00040 Monteporzio Catone, Italy\\
$^3$Space Telescope Science Institute, 3700 San Martin Drive, Baltimore, MD 21218, USA\\
$^4$Dipartimento di Fisica, Universit\`a di Roma ``La Sapienza", p.le A. Moro 2, I-00185 Roma, Italy\\ 
$^5$Astronomy Department, Columbia University, 550 West 120th Street, New York, NY 10027-6601, USA\\
$^6$Osservatorio Astronomico di Brera, INAF, via Brera 28, I-20121 Milano, Italy\\
$^7$AIM CEA/Irfu/Service d'Astrophysique, Orme des Merisiers, F-91191 Gif-sur-Yvette, France
}
\date{Accepted 2013 May 15. Received 2013 May 14; in
original form 2013 April 17} \pagerange{\pageref{firstpage}--\pageref{lastpage}} \pubyear{2013}

\def\LaTeX{L\kern-.36em\raise.3ex\hbox{a}\kern-.15em
    T\kern-.1667em\lower.7ex\hbox{E}\kern-.125emX}

\def\cxo {\emph{Chandra}}
\def\swift {\emph{Swift}}

\def\igr {\emph{INTEGRAL}}
\def\sax {\emph{BeppoSAX}}

\def\rst {\emph{ROSAT}}

\def\src {1RXS\,J2253}
\def\flux {\mbox{erg cm$^{-2}$ s$^{-1}$}}
\def\lum {\mbox{erg s$^{-1}$}}
\def\nh {$N_{\rm H}$}

\begin{document}

\label{firstpage}
\maketitle
\begin{abstract}
We report on the discovery of pulsations at a period of $\sim$47 s in the persistent X-ray source 1RXS\,J225352.8+624354 (\src) using five \cxo\ observations performed in 2009. The signal was also detected in \swift\ and \rst\ data, allowing us to infer over a 16-yr baseline an average, long-term period increasing rate of $\approx$17 ms per year and therefore to confirm the signal as the spin period of an accreting, spinning-down neutron star. The pulse profile of \src\ ($\sim$50--60\% pulsed fraction) is complex and energy independent (within the statistical uncertainties). The 1--10 keV \cxo\ spectra are well fit by an absorbed  power-law model with $\Gamma \sim 1.4$ and observed flux of (2--$5)\times10^{-12}$ \flux. The source was also detected by \igr\ in the 17--60 keV band at a persistent flux of $\sim$$6\times10^{-12}$ \flux, implying a spectral cut off around 15 keV. We also carried out optical spectroscopic follow-up observations of the  2MASS counterpart at the Nordic Optical Telescope. This made it possible to first classify the companion of \src\ as a B0-1III-Ve (most likely a B1Ve) star at a distance of about 4--5 kpc (favouring an association with the Perseus arm of the Galaxy). The latter finding implies an X-ray luminosity of $\sim$$3\times10^{34}$ \lum, suggesting that \src\ is a new member of the sub-class of low-luminosity long-orbital-period persistent Be/X-ray pulsars in a wide and circular orbit (such as X Persei).
\end{abstract}

\section{Introduction}
Every time a new X-ray mission is launched, many serendipitous X-ray sources of unknown nature are usually discovered. Among them, some may suddenly become objects of interest because they display outbursts/flares or are recognised to be the X-ray counterpart of sources discovered at other wavelengths, but most of them, especially the faintest ones, can remain unidentified for years. 
Sometimes they are `re-discovered' by the next X-ray missions or by systematic archival searches for specific signatures (such as flux variability, multiwavelength associations or pulsations).

The identification of detected sources with still unknown nature is a fundamental step towards the study of the different populations of Galactic and extragalactic X-ray sources, and can lead to surprising results. Examples are the discovery of RX\,J0806.3+1527, the double-degenerate ultracompact binary with the shortest known orbital period (5.4 minutes; \citealt{ipc99,israel02short}; \citealt*{ramsay02}), or the realisation of the existence of a potentially large population of `dormant' magnetars (e.g. \citealt{rea11} and references therein). This was also the case of the supergiant fast X-ray transients (SFXTs), a class of hard X-ray transients discovered by the \igr\ satellite  \citep{sguera05short}; in several cases in fact, SFXTs were found to be associated with objects listed in catalogues of faint soft-X-ray sources from past missions (mostly \emph{ASCA} and \sax).

The discovery of coherent X-ray pulsations, in particular, is a key element to understand the nature of a source. The \emph{Chandra ACIS Timing Survey at Brera And Rome astronomical observatories} (CATS@BAR)  is a project aimed at the exploitation of Advanced CCD Imaging Spectrometer (ACIS; \citealt{garmire03}) archival data from the \cxo\ mission \citep{weisskopf02}.\footnote{See http://www.mporzio.astro.it/gianluca/resultss.html for the analogous \swift\ project, \emph{Swift Automatic Timing ANAlysis of Serendipitous Sources at Brera And Roma astronomical observatories} (SATANASS@BAR).}  As of 2013 February 28, approximately 8,750 Timed-Exposure observations were retrieved (data taken with gratings and in Continuous-Clocking mode were not included in the analysis) and about 400,000 light curves were extracted. For the 85,000 light curves with more than $\sim$150 photons, Fourier power spectra were computed and analysed. These were searched for coherent or quasi-coherent signals in a systematic and automatised way by applying the detection algorithm described in \citet{israel96}. Among the about 30 new candidate X-ray pulsators showing signals with high confidence level ($>$$4.5\sigma$) found so far (Israel et al., in preparation), there is 1RXS\,J225352.8+624354 (hereafter \src), which has a period of $\sim$47 s. 

The source was discovered by \rst\ and included in the All-Sky Survey Faint Source Catalogue (\citealt{voges00short}; 1RXS\,J225352.8+624354) and in the WGA Catalog (\citealt*{white94}; 1WGA\,J2253.9+6243). \src\ was later suggested to be a low-luminosity hard X-ray binary by \citet{suchkov04}. More recently, \citet{landi12} pointed out that \src\ is the X-ray counterpart of the \igr\ source IGR\,J22534+6243 \citep{krivonos12}. The 47-s pulsation of \src\ was then discovered and reported by \citealt{halpern12} and independently (within the CATS@BAR project) by \citet{ir12}. It was also noticed that the position of \src\ is consistent with that of the infrared source 2MASS\,J22535512+6243368 \citep{halpern12,landi12}, suggesting a high-mass stellar companion. The high-mass X-ray binary (HMXB) classification was strengthened by \citet{masetti12} who performed a follow-up optical observation in 2012 June and obtained a  4000--8000 \AA\ spectrum characterised by a highly-reddened, intrinsically-blue continuum with H$\alpha$, H$\beta$ and He\textsc{i} emission lines. Here we report on the spectral and timing analysis of the archival \rst, \swift, and \cxo\ observations that serendipitously imaged the field of \src\ across $\sim$16 years. As a further step toward the classification of \src, we also observed it with the 2.5-m Nordic Optical Telescope in the Canary Islands on two nights in 2012 December.

\section{Observations and data reduction}
\begin{table*}
\begin{minipage}{11.cm}
\centering \caption{Observations used for this work.} \label{logs}
\begin{tabular}{@{}lcccc}
\hline
Instrument & Obs.ID  & Start date & Duration & Net count rate$^{a}$\\
 &  &  (YYYY-MM-DD) & (ks) & (counts s$^{-1}$)\\
\hline
\rst/PSPC-B & \textsc{rp900425n00} & 1992-07-29 & 4.3 & $(1.1\pm0.2)\times10^{-2}$\\
\rst/PSPC-B & \textsc{rp900425a01} & 1992-12-27 & 6.4 & $(1.7\pm0.3)\times10^{-2}$\\
\rst/PSPC-B & \textsc{rp500321n00} & 1993-06-18 & 18.5 & $(1.17\pm0.08)\times10^{-2}$\\
\rst/PSPC-B & \textsc{rp500322n00} & 1993-06-20 & 15.7 & $(0.8\pm0.3)\times10^{-2}$\\
\swift/XRT & 00206257000 & 2006-04-21 & 25.0 & $(4.2\pm0.1)\times10^{-2}$\\
\swift/XRT & 00206257001 & 2006-04-21 & 48.0 & $(3.06\pm0.08)\times10^{-2}$\\
\cxo/ACIS-S & 9920 & 2009-04-16 & 27.7 & $(6.6\pm0.2)\times10^{-2}$ \\
\cxo/ACIS-I & 10811 & 2009-04-28 & 24.4 & $(6.5\pm0.2)\times10^{-2}$\\
\cxo/ACIS-I & 10812 & 2009-05-03 & 24.8 & $(6.0\pm0.2)\times10^{-2}$\\
\cxo/ACIS-S & 10810 & 2009-05-07 & 22.8 & $(9.1\pm0.2)\times10^{-2}$\\
\cxo/ACIS-I & 9919  & 2009-05-08 & 22.5 & $(5.8\pm0.2)\times10^{-2}$\\
\hline
\end{tabular}
\begin{list}{}{}
\item[$^{a}$] Observed count rates (not corrected for PSF and effective area effects) in the energy ranges: 1--2.4 keV for \rst\ and 1--10 keV for \swift\ and \cxo.
\end{list}
\end{minipage}
\end{table*}

\subsection{\cxo}
\cxo\ imaged the position of \src\ five times in 2009 (see Table~\ref{logs}) in a campaign devoted to a portion of the Cepheus OB3 molecular cloud (see \citealt{allen12short}). The data were acquired with the ACIS instrument in Very Faint imaging (Timed Exposure) mode (time resolution: $\sim$3.2 s).

The data were reprocessed with the Chandra Interactive Analysis of Observations software (\textsc{ciao}, version 4.4) using the calibration files available in the \cxo\ \textsc{caldb} 4.4.8 database. The scientific products were extracted following standard procedures, adopting extraction regions of $\sim$6--25 arcsec radii (depending on the off-axis angle, around 15 arcmin in most pointings) for the source counts. In particular the spectra, the spectral redistribution matrices and the ancillary response files were created using \textsc{specextract}. For the timing analysis, we applied the Solar system barycentre correction to the photon arrival times with \textsc{axbary}.

\subsection{\swift}
\swift\ serendipitously observed \src\ in two contiguous observations carried out to follow the afterglow of the gamma-ray burst GRB\,060421 \citep{gbb06short}. The net exposure of the X-ray Telescope (XRT; \citealt{burrows05short}) data was 73.0 ks in photon counting (PC; full imaging, time resolution: $\sim$2.5 s) mode and 1.8 ks in windowed timing (WT; one-dimensional strip readout, time resolution: $\sim$1.8 ms) mode; since the observation was pointed at GRB\,060421, only the PC data can be used to study \src\ (see Table~\ref{logs}). The Ultraviolet/Optical Telescope (UVOT; \citealt{roming05short}) observed the field simultaneously with its optical and ultraviolet (UV) filters (see Table~\ref{uvot}).

The \swift\ data were processed and screened with standard procedures and quality cuts using \textsc{ftools} in the \textsc{HEAsoft} (version 6.12) software package and the calibration files in the 2012-04-02 \textsc{caldb} release. The XRT photon arrival times were corrected to the Solar system barycenter with \textsc{barycorr}. The XRT source counts were extracted within a 20-pixel radius (one XRT
pixel corresponds to about 2.36 arcsec). For the X-ray spectral fitting, the ancillary response files were generated with \textsc{xrtmkarf} accounting for different extraction regions, vignetting, point-spread function corrections, and dead, hot or warm pixels \citep{moretti05short}. The UVOT photometry\footnote{See \citet{poole08short} for an overview of the UVOT photometric system and \citet{breeveld11} for the most updated zero-points and count rate to flux conversion factors.} was performed with the \textsc{uvotsource} task, which calculates the magnitudes through aperture photometry. 

\subsection{\rst}
The position of \src\ occurred within the \rst\ Position Sensitive Proportional Counter (PSPC; \citealt{pfeffermann87short}) field of view in four pointings carried out between July 1992 and June 1993. However, owing to large off-axis angles and/or short duration, in three out of the four observations the exposures provide too few photons for meaningful analyses. The only observation useful for timing and spectral analysis was carried out in June 1993 (obs. ID \textsc{rp500321n00}, see Table~\ref{logs}) for an effective exposure time of $\sim$18 ks. 

The event lists and spectra for \src\ and the background were extracted from circles of $\sim$1 arcmin radius. The solar system barycenter correction to the photon arrival times was applied with the \textsc{ftools} tasks \textsc{bct} and \textsc{abc}. For the spectroscopy, we used the spectral redistribution matrix \textsc{pspcb\_gain2\_256.rmf}, while the ancillary response file was generated with \textsc{pcarf}.

\subsection{Nordic Optical Telescope}

\src\ was observed at the 2.5-m Nordic Optical Telescope (NOT) equipped with the Andalucia Faint Object Spectrograph and Camera, (ALFOSC) on 2012 December 26 and 30. We used the grism N.7 and slit of 0.5\,arcsec, covering the wavelength range $\sim$3800--6800 \AA\, at a resolution of 4.4\,\AA\, or $\sim$237\,km s$^{-1}$.\footnote{As measured on the sky emission lines.} We obtained four spectra per night, each of 900\,s exposure time. The observations had attached arc-lamp exposure and flat field.

The data were reduced using \textsc{iraf} \citep{tody93} packages and tasks and following standard procedures. However, we did not apply the flat field correction due to the fact that the flats were insufficient in number and, in one case, count level. Hence, we did not correct for the pixel-to-pixel variation, nor we removed the fringing at $\lambda\geq6300$\,\AA. However, we removed the spectrograph signatures (low frequency variations) during the flux calibration step: the spectro-photometric standard stars were observed with exactly the same set up. In particular we observed both  BD\,+17$^{\circ}$4708 and Feige\,110, during each night. Fluxes are not absolute as both the target object and the spectrophotometric standard star were affected by slit losses which we could not estimate and correct for.

\section{Timing analysis and results}\label{timing}

The inspection of the X-ray light curves showed evidence of moderate variability on a time-scale of a few ks. The root-mean-square (rms) fractional variation (defined as the rms variation normalised by the average count rate) was $52\pm5\%$ in observation \swift/7000, and $47\pm6\%$ in \swift/7001 (measured in 1--10 keV light curves, bin size of 500 s). In the \cxo\ observations (1--10 keV,  500 s bin size) it was, in chronological order, $22\pm4\%$ (obs. 9920), $34\pm4\%$ (10811), $29\pm4\%$ (10812), $33\pm4\%$ (10810), and $23\pm4\%$ (9919). For the \rst\ observation 500321 we could only place an upper limit of $\sim$40\% ($3\sigma$ c.l.) for bin sizes from 0.5 to 5 ks (1--2.4 keV). 

\begin{figure}
\centering
\resizebox{\hsize}{!}{\includegraphics[angle=-90]{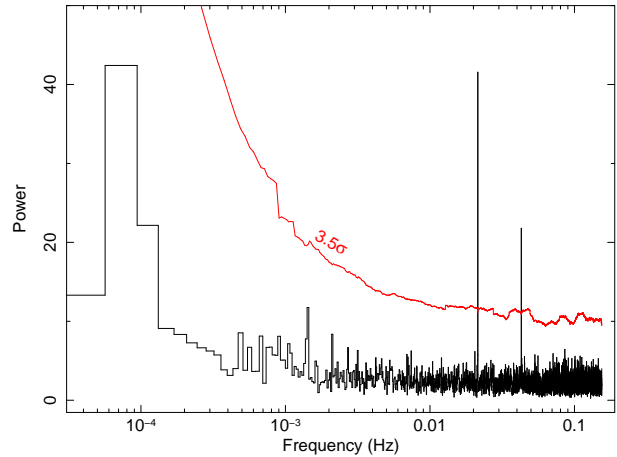}}
\caption{\label{powerspec} Fourier power spectrum of the whole \cxo\ dataset of \src\ (0.5--10 keV). The red stepped line corresponds to 3.5$\sigma$ confidence level threshold for potential signals (computed taking into accounts the number of trial equal to the number of frequency bin of the spectrum). The two peaks above the threshold are the fundamental and the second harmonics of the 46.7-s signal.}
\end{figure}

As anticipated, $\sim$46.7-s coherent pulsations from \src\ were found within the CATS@BAR project using the whole \cxo\ dataset. Fig.\,\ref{powerspec} shows the discovery periodogram where two peaks, corresponding to the fundamental ($\nu_1=1/P\simeq0.0214$ Hz) and the second ($\nu_2=1/(2P)$) harmonics, stand well above the significance threshold. For the analysis of the coherent period that follows, the photon arrival times were transformed to Barycentric Dynamical Time (TDB) using the 2MASS coordinates of the optical/infrared counterpart listed in Table~\ref{tab:ephemeris}.
 
\begin{figure}
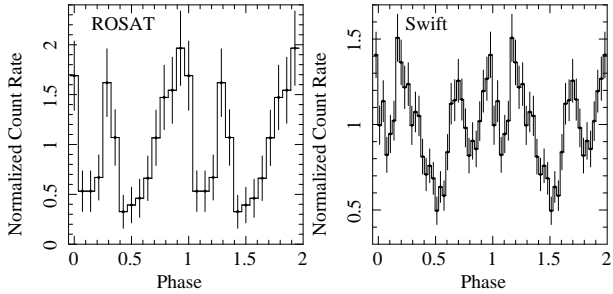

\centering
\resizebox{\hsize}{!}{\includegraphics[angle=-90]{fig2a.ps}\hspace{-1.2cm}\includegraphics[angle=-90]{fig2b.ps}}
\caption{\label{rstprofile}\rst/PSPC (0.1--2.4 keV) and \swift/XRT (1--10 keV) normalised pulse profiles of \src. Both profiles have been aligned so that the minimum phase coincides with that of the \cxo\ minimum in Fig.~\ref{fig:chandrapulse}.}
\end{figure}

The signal is easily detectable also in the 2006 \swift\ dataset (since the two observations were contiguous, we treated them as a single one), where we measured by the $Z^2_2$ test (e.g. \citealt{buccheri83short}), which optimises the pulsed power, a period of 46.6145(5) s. The pulsed fraction (defined as $(M-m)/(M+m)$, where $M$ and $m$ are the observed background-subtracted count rates at the peak and at the minimum, respectively) of the folded profile (Fig.\,\ref{rstprofile}) is $45\pm5$ percent in the 1--10 keV range.

In the case of the \rst/PSPC data, considering the $\sim$13 years elapsed between \rst\ and \swift\ observations, we carried out a period search in an interval set by a period derivative $\pm$$\dot{P}$, assuming a maximum spin-up/down intensity $|\dot{P}| = 5\times10^{-8}$  s s$^{-1}$. A significant  peak (6.4$\sigma$ c.l. in about 10$^3$ trial periods) was found in a $Z^2_2$ periodogram at a best period of 46.406(5) s (see the folded profile in Fig.\,\ref{rstprofile}). The corresponding pulsed fraction is $65\pm11$ percent (0.1--2.4 keV).

\begin{table}
\centering \caption{\cxo\ timing of \src.} \label{tab:timingdata}
\begin{tabular}{@{}lcccc}
\hline
Obs.ID & Start Epoch & Period & $Z^2_2$ & Pulsed fraction\\
      & MJD         & (s)    &  & (\%)\\
\hline
9920   & 54937.446  &  46.6784(37)  & 94.7 & $55\pm7$ \\
10811  & 54949.292  &  46.6718(44)  & 94.7 & $60\pm6$ \\
10812  & 54954.072  &  46.6703(37)  & 90.4 & $48\pm7$ \\
10810  & 54958.855  &  46.6701(37)  & 96.2 & $47\pm6$ \\
9919   & 54959.143  &  46.6646(41)  & 79.8 & $47\pm7$ \\
\hline
\end{tabular}
\end{table}

\begin{figure}
\resizebox{\hsize}{!}{\includegraphics[angle=-90]{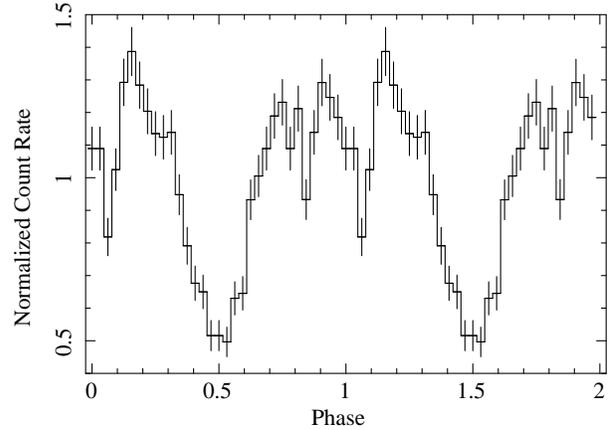}}
\caption{Normalised pulse profiles of \src\ in the 1--10~keV band from \cxo, obtained by folding all data with the ephemeris in Table~\ref{tab:ephemeris}. The TDB epoch of pulse minimum (phase 0.5) is MJD 54954.00050.
}
\label{fig:chandrapulse}
\end{figure}

For each ACIS observation listed in Table~\ref{logs} we extracted source photons in the 1--10~keV band using an aperture of radius $10^{\prime\prime}$ or $20^{\prime\prime}$ as appropriate. We used the $Z^2_2$ test to measure the periods in the individual observations (Table~\ref{tab:timingdata}), and to bootstrap a coherent ephemeris that links all of them. Beginning with the last two observations, which were contiguous over May 7--8 (Obs.IDs 10810 and 9919), it was possible to work backwards in time and add Obs.ID 10812 on May 3 and Obs.ID 10811 on April 28 to the ephemeris, refining the period as each additional observation was included.  It was not possible to extrapolate further back to Obs.ID 9920 on April 16 using a linear ephemeris (constant period), as the phase of that observation deviated by $\approx$0.6 cycles from the predicted phase.  However, a fit including a period derivative accurately predicted its phase, and resulted in the quadratic ephemeris given in Table~\ref{tab:ephemeris}.  Figure~\ref{fig:chandrapulse} shows the pulse profiles folded according to this ephemeris.

\begin{table}
\centering \caption{\cxo\ ephemeris of \src.} \label{tab:ephemeris}
\begin{tabular}{@{}lr}
\hline
Parameter & Value \\
\hline
R.A. (J2000.0)$^{a}$ & $22^{\rm h}\,53^{\rm m}\,55\fs12$ \\
Decl. (J2000.0)$^{a}$ & $+62^{\circ}\,43^{\prime}\,36\farcs8$ \\
Epoch of ephemeris (MJD TDB)$^{b}$ & 54954.00050 \\
Valid range of dates (MJD)                    & 54937--54959 \\
Frequency, $f$ (Hz)                           & 0.02142536(2) \\
\medskip
Frequency derivative, $\dot f$ (Hz s$^{-1}$)  & $-6.1(4)\times10^{-13}$ \\
Period, $P$ (s)                                   & 46.67366(4) \\ 
Period derivative, $\dot P$ (s s$^{-1}$)                   & $1.33(9)\times10^{-9}$\\
\hline
\end{tabular}
\begin{list}{}{}
\item[$^{a}$]2MASS coordinates of optical/infrared counterpart.
\item[$^{b}$]Epoch of pulse minimum (phase 0.5) in Fig.\,\ref{fig:chandrapulse}.
\end{list}
\end{table}

\begin{figure}
\resizebox{\hsize}{!}{\includegraphics[angle=0]{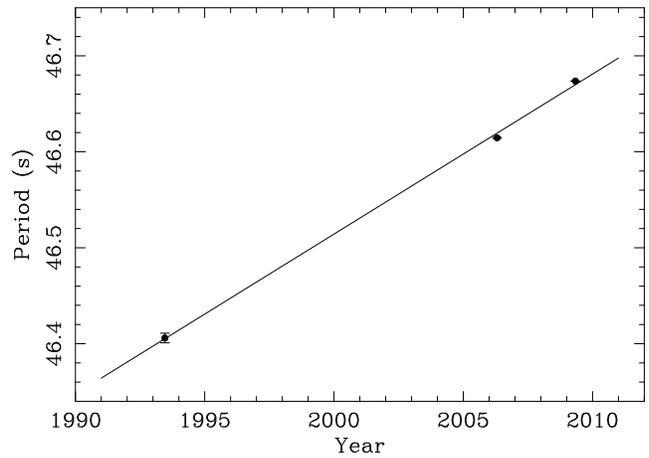}}
\caption{Long term spin-down of \src\ from \rst, \swift, and \cxo. The systematic error on $P$ due to orbital motion is $\approx$0.02~s (Eq.~\ref{eq:2}), which is larger than the measurement error on each point. Therefore, a simple least-squares fit is employed, yielding a mean $\dot P = 5.3\times10^{-10}$ s s$^{-1}$. }\label{fig:spindown}
\end{figure}

The fit period derivative, $\dot P=1.3\times10^{-9}$ s s$^{-1}$, may have contributions from both secular spin-down and orbital acceleration, as we show here. The long-term spin-down rate is $\approx$$5.3\times10^{-10}$ s s$^{-1}$ (Fig.~\ref{fig:spindown}). This is only $\approx$$40\%$ of the $\dot P$ required to fit the series of \cxo\ observations, which means that orbital acceleration contributes to the \cxo\ timing over 22 days. The kinematic effect may be estimated assuming $M_{\rm x}=1.4\,M_{\odot}$ for the neutron star in a circular orbit around a Be star companion (see Sect.\,\ref{not}) of mass $M_{\rm c}=20\,M_{\odot}$. The contribution of acceleration to $\dot P$ is
\begin{eqnarray}
\dot P_{\rm k} & = & 1.4\times10^{-8}
\left({P_{\rm orb} \over 100\,{\rm d}}\right)^{-4/3}
\left({M_{\rm c} \over 20\,M_{\odot}}\right)^{1/3}\nonumber\\
&&\left({P \over 46.7\,{\rm s}}\right)\
(1+q)^{-2/3}\
{\rm sin}\,i\ {\rm sin}\,\phi,\label{eq:1}
\end{eqnarray}
where $q=M_{\rm x}/M_{\rm c}$ is the mass ratio, $i$ is the inclination of the binary, and $\phi$ is the orbital phase of the companion. Evidently, the orbital period $P_{\rm orb}$ is significantly longer than the 22 day span of the \cxo\ observations because the observed $\dot P=1.3\times10^{-9}$ s s$^{-1}$ is small and constant to within $\sim$$10\%$ over this interval. Eq.~\ref{eq:1} also implies that $P_{\rm orb}<800$~days in order to provide the required acceleration. While orbital acceleration may account for the \cxo\ measured  $\dot P$, the monotonic increase in spin period between \rst, \swift, and \cxo\ timings over 16 years cannot be due to orbital motion, since the total observed range in period is $\Delta P \simeq 0.27$~s, while the kinematic effect provides only
\begin{eqnarray}
\Delta P_{\mathrm{k}} & = &  -0.019
\left({P_{\rm orb} \over 100\,{\rm d}}\right)^{-1/3}
\left({M_{\rm c} \over 20\,M_{\odot}}\right)^{1/3} \nonumber\\
&& \left({P \over 46.7\,{\rm s}}\right)\
(1+q)^{-2/3}\
{\rm sin}\,i\ {\rm cos}\,\phi\ \ {\rm s}.\label{eq:2}
\end{eqnarray}

\begin{figure}
\resizebox{\hsize}{!}{\includegraphics[angle=-90]{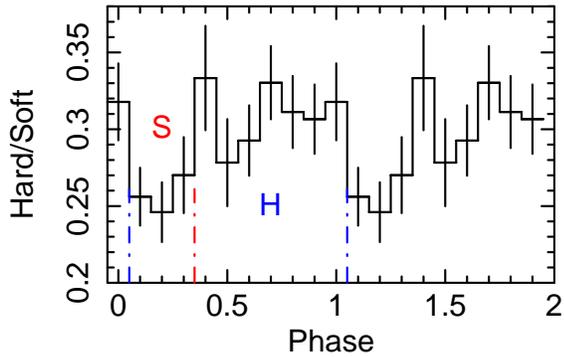}}
\caption{\label{cxohr} \cxo\ hardness ratio as function of the phase. The ratio was computed selecting soft photons in the 1--4.5 keV and hard photons in the 4.5--10 keV range. Phases are aligned with those of Fig.~\ref{fig:chandrapulse}. The vertical lines and the labels indicate the phase intervals used for the phase-resolved spectroscopy (see text).}
\end{figure}

The \rst\ pulse shape, which is shown in Fig.\,\ref{rstprofile}, is dominated by two asymmetric peaks which are energy independent within the PSPC energy range (0.1--2.4 keV). In Fig.~\ref{fig:chandrapulse} we show the epoch-folded \cxo\ pulse profiles obtained using the ephemeris of Table\,\ref{tab:ephemeris}. To assess the significance of the pulse shape variations through the high-count-statistics five \cxo\ data sets (Fig.~\ref{fig:chandrapulse}), we compared each of the folded light curves with all the others by using a bidimensional Kolmogorov-Smirnov test (e.g. \citealt{press92}). The results show that all profiles are consistent with coming from the same distribution; the most significantly different pair of profiles are those of observations 10811 and 10810, which differs at a $\sim$1.5$\sigma$ confidence level. We compared in the same way the total \cxo\ and \swift\ (Fig.\,\ref{rstprofile}) profiles. Taking into account the unknown relative phase alignment, the probability that they come from the same underlying distribution is about 6.1 per cent. The pulse profiles do not show dramatic shape variations with energy, but a hardness-ratio analysis suggests some spectral variability along the spin phase (see  Fig.~\ref{cxohr}). In fact the emission appears to be slightly harder during the minimum-rise part of the profile than at the decline phases.

\section{X-ray spectral analysis and results}

Spectral fitting was carried out with \textsc{xspec} v.12.7 \citep{arnaud96}. For each observation we extracted an average spectrum. A simple power-law model (modified for the interstellar absorption) provides an acceptable fit for all the data sets (see Table~\ref{fits} for a summary of the spectral analysis). The average measured absorbing column corresponds to $\sim$$1.9\times10^{22}$ cm$^{-2}$ and the power law is rather hard, with an average photon index $\Gamma\sim1.4$; in the \swift\ and \cxo\ observations, the observed flux varies from $\sim$$2.4\times10^{-12}$ to $\sim$$4.5\times10^{-12}$ \flux.

Inspection of the energy-resolved light curves did not reveal neat spectral variations correlated with the source intensity. On the other hand, the hardness ratio analysis presented in Sect.\,\ref{timing} suggests some spectral variations with the spin phase. For each \cxo\ observation we extracted `soft' and `hard' spectra in two intervals chosen following the hardness ratio variations (see Fig.~\ref{cxohr}). A simultaneous fit of the spectra with an absorbed power-law model ($\chi^2_\nu=1.04$ for 327 dof) confirms a moderate hardening in the minimum-rise part of the profile, with an average power-law photon index $\Gamma_{\mathrm{H}} = 1.26\pm0.06$ to be compared with  $\Gamma_{\mathrm{S}} = 1.4\pm0.1$ during the decline phases. We also produced a bidimensional histogram of the distribution of counts versus energy and phase to look for spectral features in the X-ray emission of \src, but none was found.

\begin{figure}
\centering
\resizebox{\hsize}{!}{\includegraphics[angle=-90]{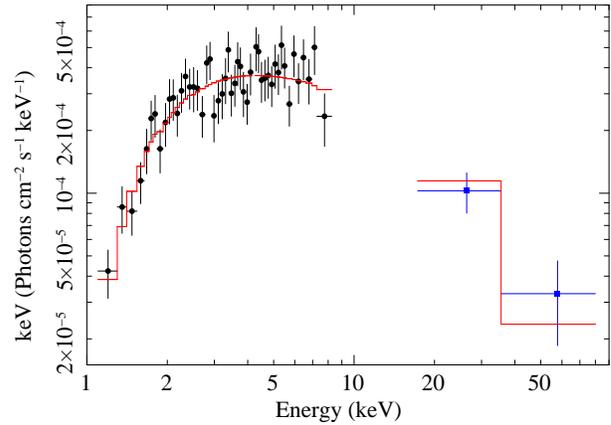}}
\caption{\label{bbspec} Broad-band soft and hard X unfolded model and (absorbed cut-off power law) spectrum ($EF(E)$) of \src. For clarity, we show only the \swift\ (black dots) and \igr\ (blue squares) data.}
\end{figure}

The hard-X-ray counterpart to \src\ has been identified with \igr\ (IGR\,J22534+6243; \citealt{krivonos12,landi12}) in the IBIS 9-year Galactic Hard-X-Ray Survey (Perseus Arm). The reported source flux in the 17--60 keV energy band is $(6\pm1)\times10^{-12}$ \flux. The extrapolation in this hard-X-ray band of the power-law spectrum that describes the 1--10 keV data overestimates the \igr\ flux by a factor $\sim$2.  We built an IBIS/ISGRI \citep{ubertini03short,lebrun03short} spectrum starting from the count rates in the 17--80 keV energy bands published by \citet{krivonos12} and rebinned the ISGRI response matrix in order to cope with the energy bands. A good simultaneous fit of the average 1--10 keV and \igr\ data ($\chi^2_\nu=1.01$ for 338 dof) can be obtained by using an absorbed cut-off power law (see Fig\,\ref{bbspec}). The resulting spectral parameters are absorption $N_{\mathrm{H}}=(1.6\pm0.1)\times10^{22}$ cm$^{-2}$, photon index $\Gamma=1.0\pm0.1$, cut-off energy $E_\mathrm{C}=15\pm4$ keV, and 1--60 keV unabsorbed flux of $\approx$$1.2\times10^{-11}$ \flux\ (the 17--60 keV flux is $\approx$$5\times10^{-12}$ \flux). This spectrum (slope and high energy cut-off) is typical of X-ray accreting pulsars (e.g. \citealt{white83}).

\begin{table*}
\centering 
\begin{minipage}{11.5cm}
\caption{Spectral analysis of \src. Errors are at a 1$\sigma$ confidence level for a single parameter of interest.} \label{fits}
\begin{tabular}{@{}cccccc}
\hline
Observation  & \nh & $\Gamma$ & Observed flux$^a$ & Unabsorbed flux$^a$ & $\chi^2_\nu$ (dof)\\
& ($10^{22}$ cm$^{-2}$) & & \multicolumn{2}{c}{($10^{-12}$ \flux)} &\\
\hline
\rst\ 500321 &  $1.8^{+1.1}_{-0.7}$ & $3^{+3}_{-2}$ & $0.5\pm0.2$ & $4^{+90}_{-3}$ & 0.60 (12)\\
\swift\ 7000 & $1.7\pm0.2$& $1.2^{+0.2}_{-0.1}$ & $4.5^{+0.3}_{-0.2}$ & $6.2\pm0.3$ & 1.04 (48)\\
\swift\ 7001 & $2.0\pm0.2$& $1.5\pm0.1$ & $3.0\pm0.1$ & $4.7\pm0.3$ & 1.12 (67)\\
\cxo\ 9920 &  $1.6^{+0.1}_{-0.2}$& $1.2\pm0.1$ & $2.7^{+0.1}_{-0.2}$ & $3.6\pm0.1$ & 0.91 (88)\\
\cxo\ 10811 &  $1.8\pm0.2$& $1.3\pm0.1$ & $3.3\pm0.2$ & $4.7\pm0.2$ & 0.79 (68)\\
\cxo\ 10812 &  $2.0\pm0.2$& $1.4^{+0.1}_{-0.2}$ & $2.4\pm0.1$ & $3.6\pm0.2$ & 0.84 (63)\\
\cxo\ 10810 &  $1.9^{+0.2}_{-0.1}$& $1.4\pm0.1$ & $3.4^{+0.1}_{-0.2}$  & $5.0\pm0.2$ & 0.94 (89)\\
\cxo\ 9919 &  $1.8^{+0.3}_{-0.2}$& $1.2^{+0.2}_{-0.1}$ & $2.7^{+0.1}_{-0.2}$  & $3.7^{+0.1}_{-0.2}$ & 1.24 (56)\\
\hline
\end{tabular}
\begin{list}{}{}
\item[$^{a}$] In the 0.5--10 keV energy range.
\end{list}
\end{minipage}
\end{table*}

\section{Optical and ultraviolet analysis and results}

As observed by \citet{landi12} and \citet{halpern12}, the X-ray position of \src\ is consistent with that of the Two-Micron All-Sky Survey (2MASS; \citealt{skrutskie06short}) source 2MASS\,22535512+6243368 (see Table~\ref{uvot} for its magnitudes). The next closest source to \src, 2MASS\,22535500+6243412, lies $\sim$4.4 arcsec away.

\begin{figure}
\centering
\resizebox{.9\hsize}{!}{\includegraphics[angle=0]{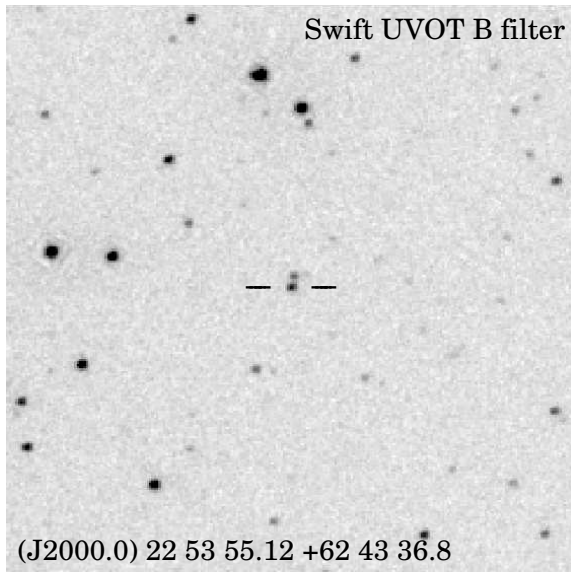}}
\caption{\label{findchart}\swift/UVOT $B$-band image ($4'\times4'$) centred on 2MASS\,22535512+6243368 (marked with solid lines).}
\end{figure}

Since the 2MASS sources are slightly blended in the UVOT images (see Fig.\,\ref{findchart}), we used for the photometry a small aperture radius of 2 arcsec (no attempt was made to correct for possible residual contamination from 2MASS\,22535500+6243412). The source is detected with high confidence in all the UVOT filters but the $uvm2$, in which the statistical significance of the source is only at a 3.2$\sigma$ confidence level. The average  Vega UVOT magnitudes, calculated from the stacked images using the \textsc{uvotsource} tool, are reported in Table~\ref{uvot}.

\begin{table}
\centering \caption{UVOT magnitudes, not corrected for the extinction. We also give the catalogued 2MASS infrared magnitudes} \label{uvot}
\begin{tabular}{@{}lcc}
\hline
Filter & Magnitude & Exposure\\
 &  & (ks)\\
\hline
$V$ & $15.78\pm0.09$ & 7.5 \\
$B$ & $17.38\pm0.11$ & 6.4 \\
$U$ & $17.64\pm0.12$ & 9.5 \\
$uvw1$ & $19.24\pm0.13$ & 15.1 \\
$uvm2$ & $22.34\pm0.38$ & 17.2 \\
\medskip
$uvw2$ & $20.54\pm0.16$ & 19.2 \\
$J$ & $11.64\pm0.02$ & --\\
$H$ & $10.96\pm0.02$ & --\\
$K$ & $10.46\pm0.03$ & --\\
\hline
\end{tabular}
\end{table}

\subsection{Spectral type of the stellar companion}\label{not}
\begin{figure*}
\centering
\resizebox{\hsize}{!}{\includegraphics[angle=270]{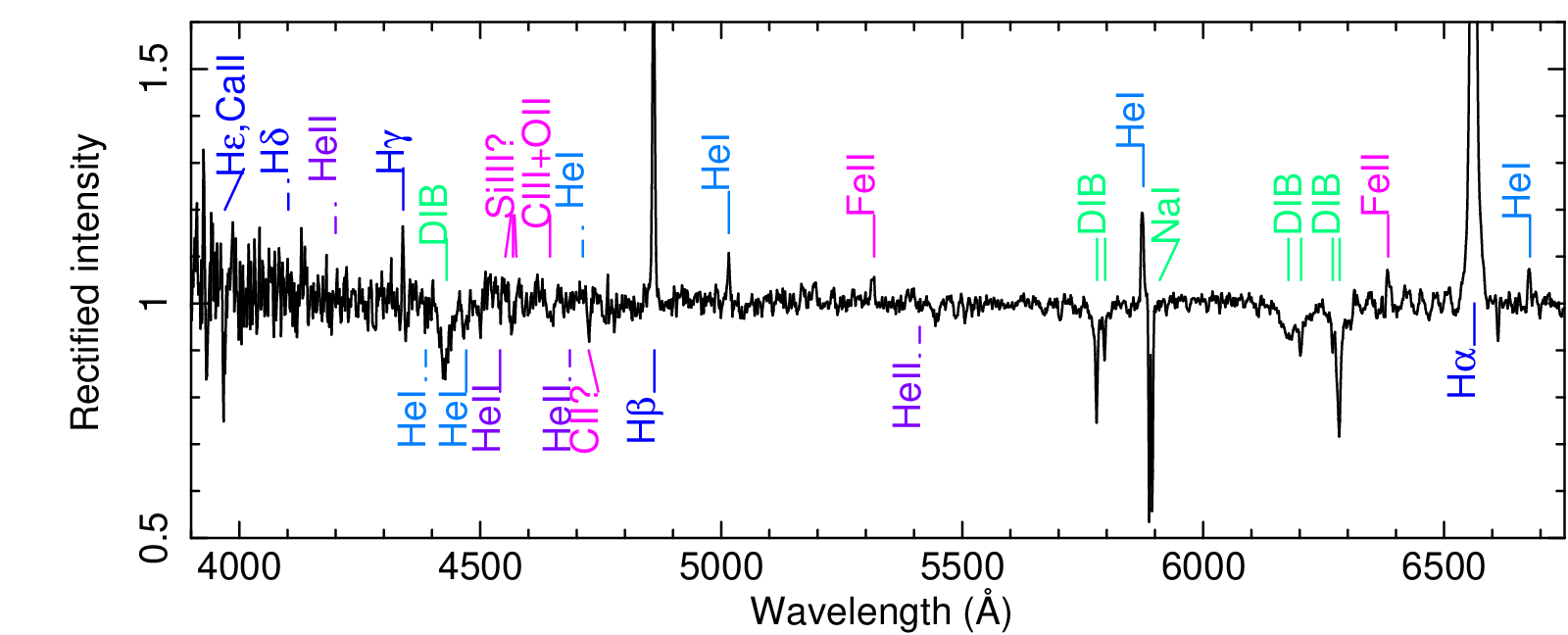}}
\caption{\label{fig:specnot}NOT/ALFOSC medium resolution (4.4 \AA; 3800--7000 \AA) rectified spectrum obtained for the optical counterpart of \src. The main identified lines are marked with solid lines and labelled. Dot-dashed lines mark further lines often used for spectral classification but absent in the spectrum.}
\end{figure*}

The optical spectrum shows the presence of H$\alpha$ (equivalent width, $EW \sim-38$ \AA), H$\beta$ ($EW \sim-5$ \AA) and H$\gamma$ ($EW \sim-1$ \AA) in emission (H$\delta$ seems to be filled in while H$\epsilon$ is in absorption) together with He\textsc{i} emission lines at 6678 \AA, 5875 \AA\ and 5016 \AA\ (the He\textsc{i} at 4713 \AA\ is likely filled in while He\textsc{i} at 4471 \AA\ is in absorption)  and a number of metallic lines typical of an early B-type star of low luminosity (Fig. \ref{fig:specnot}). More precisely, the absence of (or very weak) He\textsc{ii} lines indicates an spectral type later than B1. He\textsc{i} lines dominate the spectrum (together with the hydrogen lines), indicating a B0 to B2 star. However, the presence of some amount of Si\textsc{iii} 4552--68 \AA\ (the identification of this lines is unclear since we see only one of the triplet transition) and the carbon blends (C\textsc{iii} 4650 \AA) favours the B0--B1 spectral type. On the other hand, the weakness/absence of the oxygen and silicon lines points toward a main-sequence star, although the strength of C\textsc{iii} 4650 \AA\ might be a signature of a giant companion. We tentatively conclude that the optical counterpart to \src\ is likely a B0V--B1V star (more likely a B1V), although a more luminous companion cannot be ruled out by present data. 

Assuming typical colours of a B1V star, $(B-V)_0=-0.23$ \citep{wegner94}, and comparing with the observed colour $(B-V)_{\rm obs} = 1.49$--1.7 (Table\,\ref{uvot}), we derived an excess colour of $E(B-V)=1.72$--1.93. Assuming an absolute magnitude of $M_V=-3.2$ \citep{gray09}, we estimate the distance to be $\sim$4--5 kpc (which is consistent with the Perseus arm; e.g. \citealt{russeil03}). Similar reasoning applies for a B1I and a B1II spectral class and imply distances in the $\sim$14--19 kpc and  $\sim$9--12 kpc ranges, respectively. The Galaxy edge of about 10 kpc in the direction of \src\ rules out the possibility that the companion star is a B1I, while is marginally in agreement with a B1II spectral type (which, however,  seems to be incompatible with the apparent lack of O\textsc{ii} absorption lines at 4415--17 \AA).

\section{Discussion}
We reported here on the discovery of 47-s spin pulses in the X-ray emission of \src\ and on a multiwavelength and long-term ($\sim$16~years) study of the source which made it possible to put stringent constraints on its nature. The optical spectroscopy indicates a B1-type companion, very likely a main sequence star at a distance of $\sim$4--5 kpc, implying that \src\ can be classified as a massive X-ray binary located in the Perseus arm of our Galaxy. This estimated range for the pulsar distance translates into an X-ray luminosity of (2.3--$3.6)\times10^{34}$ \lum\ assuming an average flux of $1.2\times10^{-11}$ \flux\ (1--60 keV, corrected for the absorption). The X-ray emission at a level of $\approx$$10^{34}$ \lum\ in observations spaced by several years, strongly suggests that \src\ is a member of the  sub-class of Be X-ray pulsars where the persistent and low X-ray luminosity is driven by the accretion onto the neutron star of wind material from the massive companion in a wide (orbital period, $P_{\rm orb}$, longer than $\sim$30~days) and nearly circular ($e<0.2$) orbit. These sources, the prototype and most luminous of which is X Persei \citep{white76,delgado01,lapalombara07}, were recognised by \citet{pfahl02} as a new sub-class of Be/X-ray binaries (XRBs), characterised by a smaller natal kick compared to classical Be/XRBs with more eccentric orbits.

The long-term pulse period changes observed in \src\ indicate that in the years 1993--2009 the neutron star slowed down at an average rate of $\sim$$5.3\times10^{-10}$ s s$^{-1}$; this estimate is based, however, on determinations of the period in only three epochs and no information is available on the spin behaviour of \src\ on shorter time-scales (a few months or less). As discussed in Sect.\,\ref{timing}, the timing analysis of \src\ also shows that the orbital period of the system has to be longer than $\approx$20 days (and shorter than $\approx$800 days). In Be/XRBs, the pulsar spin period is correlated with the orbital one \citep{corbet86}, although with a large observed scatter. If the same holds also for \src, given the 47-s spin period, its orbital period should be around 70--80 days \citep{corbet86}. 

The long orbital period in \src\ can explain the low X-ray luminosity, since it is produced by the accretion from the wind of the main-sequence early-type companion at a large orbital separation. Given that X-ray pulsations are detected in all the reported observations, \src\ always undergoes accretion with the material channelled onto the neutron star polar caps, even at these low X-ray luminosities. In the simpliest framework for wind-fed binaries, the accretion luminosity scales as $L_{\mathrm{X}} \propto \dot{M}_w \, a^{-2} \, v_w^{-4}$, where $\dot{M}_w$ is the wind mass loss rate, $a$ is the orbital separation and $v_w$ is the wind velocity at the neutron star orbit \citep{davidson73}. Assuming $\dot{M}_w \sim 10^{-7}$--$10^{-8}$ $M_\odot$ yr$^{-1}$ for an early-type main sequence star \citep{kudritzki00}, $v_w \sim 500$--1000 km s$^{-1}$, $a \sim (5$--$10) \times 10^{12}$ cm, an X-ray luminosity in the range $L_{\mathrm{X}}\sim10^{33}$--$10^{34}$ \lum\ can be accounted for.\footnote{If \src\ is accreting with the same mechanism  from a slower and denser wind component of the BV star, for a neutron-star magnetic field $B\la10^{12}$ G and assuming typical parameters of Be decretion discs \citep{waters89},  a low luminosity of $\sim$$10^{34}$ \lum\ can be obtained for very wide circular orbits, with $P_{\mathrm{orb}}\sim300$ days. See \citealt{delgado01} and \citealt{doroshenko12} for a discussion of this scenario for \mbox{X Persei}.} 
In this scenario, a long-term spin-down behaviour, as seen in \src, can be explained by angular momentum removal from the rotating neutron-star magnetosphere through an extended quasi-static shell formed by the accreting matter (\citealt{shakura12}; see e.g. \citealt*{davidson73,davies79,davies81,illarionov90,bisnovatyi91} for other torque models for wind accretion).

A somewhat different picture can be considered based on the presence of rotationally-dominated, quasi-Keplerian discs around Be stars (e.g. \citealt{struve31,marlborough69,waters89,hanuschik96,hhs96,hummel97,okazaki97}). Prompted by this, scenarios where the neutron star is capturing the matter from the dense and slow circumstellar  disc have been developed to account for the outbursting behaviour of some Be/XRBs (\citealt*{taam88}; \citealt{waters88,parmar89}). In particular, \citet{okazaki01} showed that the circumstellar matter can reach the neutron star only via the inner Lagrangian point (the position of which is variable along the orbit in elliptical or moderately elliptical systems), and will therefore have a low velocity relative to the neutron star. Since such a flow (which may be said to represent the `decretion-disc version' of Roche-lobe overflow) carries angular momentum, an accretion disc may be temporarily formed around the neutron star. It is therefore possible that \src\ is accreting from a similar (but more stable) structure. Depending on the orbital parameters (orbital separation, orbital period and eccentricity) and the neutron star properties (such as magnetic field and magnetic moment axis inclination with respect to the disc),  the luminosity level observed in \src\ can be accounted for and the pulsar can show both spin-up and spin-down \citep*{ghosh79,gl79,lovelace95,perna06}.

Finally we note that \src\ is associated with the hard X-ray source IGR~J22534+6243 \citep{landi12}, a faint source reported for the first time by \citet{krivonos12} in the \igr/IBIS nine-year Galactic hard X-ray survey catalog, with an average flux of $(6\pm1)\times10^{-12}$ \flux\ in the 17--60 keV energy range. To date (February 2013) the source sky position has been observed by \igr/IBIS for a net exposure time of 7.5 Ms, and no bright flares nor outbursts have ever been reported, consistently with a faint persistent source. The almost constant (within a factor of $\sim$2) X-ray luminosity (note the large uncertainty in the extrapolation of the \rst\ flux to higher energies) points to a circular, or nearly circular, orbit.

High mass X-ray binaries composed of neutron stars accreting from main-sequence companions are likely to constitute a not negligible fraction of the unidentified Galactic X-ray sources with relatively low luminosities, $L_{\mathrm{X}}$ less than $10^{35}$ \lum\ (perhaps a few percents but less than 10\%; see e.g. \citealt{laycock05,hong12}). Our findings suggest that \src\ is a new member of an elusive population of persistent low-luminosity Galactic HMXBs accreting from B main-sequence donors.
 
\section*{Acknowledgments} 
The project leading to these results has received funding from the European Union Seventh Framework Programme (FP7/2007--2013) under grant agreement No. 312430 (OPTICON). This research is based on data and software provided by the CXC (operated for NASA by SAO), the NASA's HEASARC and the ASI Science Data Center (for the \swift\ XRT Data Analysis Software, XRTDAS), and on observations made with the NOT, operated by the NOTSA in the IAC's Observatorio del Roque de los Muchachos.
We also made use of data products from the Two Micron All Sky Survey, which is a joint project of the University of Massachusetts and the IPAC/Caltech, funded by the NASA and the NSF. We thank Enrico Bozzo for useful discussions.

\bibliographystyle{mn2e}
\bibliography{biblio}

\bsp

\label{lastpage}

\end{document}